\documentclass[12pt]{article}
\usepackage{graphicx}
\usepackage{epsfig}
\usepackage{amsmath}
\usepackage{amsthm}
\usepackage{amssymb}

\usepackage[round]{natbib}
\usepackage{color}

\usepackage{units}

\newcommand{\n}[1] {\mbox{\boldmath{$#1$}}}

\DeclareMathOperator{\plim}{plim}

\newtheorem{res}{Result}
\newtheorem{lem}{Lemma}
\newtheorem{thm}{Theorem}
\newtheorem{defn}{Definition}
\newcommand{\be}{\begin{eqnarray}}
\newcommand{\ee}{\end{eqnarray}}
\newcommand{\beq}[1]{\begin{equation}\label{#1}}
\newcommand{\eeq}{\end{equation}}
\newcommand{\ba}{\begin{eqnarray*}}
\newcommand{\ea}{\end{eqnarray*}}

\title{Bayesian variable selection in high dimensional problems\\ without assumptions on prior model probabilities}
\author{J. O. Berger$^1$, G. Garc\'ia-Donato$^2$, M. A. Mart\'inez-Beneito$^3$ and V. Pe\~na$^1$\\
\footnotesize{$^1$ Duke University; $^2$ Universidad de Castilla La Mancha; $^3$ FISABIO (Valencia)}}

\date{\today}
\begin{document}
\maketitle


\begin{abstract}
We consider the problem of variable selection in linear models when $p$, the number of potential regressors, 
may exceed (and perhaps substantially) the sample size $n$ (which is possibly small).
\end{abstract}

\section{Introduction and notation}
In model selection problems the uncertainty about which model has generated the data is explicitly considered. Variable selection is a particular problem of model selection where models share a common functional form but differ in the explanatory variables that constitute the models. 
We consider the problem of variable selection in linear models when $p$, the number of potential regressors, may exceed (and perhaps substantially) the sample size $n$ (which is possibly small). See \cite{West02, JoTi09} for excellent introductions to the topic. 

Let $\n y$ be a sample of $n$ observations of the response variable and let $\n X$ be the $n\times p$ design matrix containing by columns the potential explanatory variables. As previously used in the literature we compactly express the set of all candidate models $M_\gamma\in{\cal M}$ using a binary vector $\n\gamma^t=(\gamma_1,\ldots,\gamma_p)$ where each $\gamma_i$ is zero or one indicating whether the $i$-th covariate is included or not in $M_\gamma$. Hence ${\cal M}=\{M_\gamma:\n\gamma\in \{0,1\}^p\}$ where
$$
M_\gamma:\,\, \n y=\alpha\n 1_n+\n X_\gamma\n\beta_\gamma+\n\epsilon,\,\,\,\n\epsilon\sim N(\n 0,\sigma^2\n I_n),
$$
and $\n X_\gamma$ is the $n\times k_\gamma$ sub-matrix of $\n X$ with columns defined by the 1's in $\n\gamma$ and with associated regression parameter $\n\beta_\gamma$ of dimension $k_\gamma=\sum_{i=1}^p\,\gamma_i$. We assume that if $n>k_\gamma$ then $rank(\n 1,\n X_\gamma)=k_\gamma+1$ and if $n\le k_\gamma$ then $rank(\n X_\gamma)=n$. Finally, denote $\n V_\gamma$ the $\n X_\gamma$ with the columns centered on their means (i.e. $\n V_\gamma=(\n I-\n P_n)\n X_\gamma$ where $\n P_n=\n 1_n\n 1_n^t/n$ is the orthogonal projection onto the vector space defined by the intercept). 

We denote $M_0$ the null model ($\n\gamma=\n 0$) that has $k_0=1$ regressors (just the intercept). The problem with the null model containing no regressors ($k_0=0$) is very similar and will be considered throughout the paper. In this case, it is assumed that if $n\ge k_\gamma$ then $rank(\n X_\gamma)=k_\gamma$ and if $n< k_\gamma$ then $rank(\n X_\gamma)=n$. Further $\n V_\gamma=\n X_\gamma$. In order to not duplicate all the formulas, at the price of abusing slightly notation, in what follows it should be understood that the parameter $\alpha$ does not exist when $k_0=0$. 

The formal Bayesian answer to the model selection problem is based on the posterior distribution over the model space
$$
f(\n\gamma\mid\n y)\propto B_\gamma\, f(\n\gamma)
$$
where $f(\n\gamma)$ is the prior probability of $M_\gamma$ and $B_\gamma$ is the Bayes factor (see ) of $M_\gamma$ to a fixed model here taken as $M_0$. 


$B_\gamma$ is the ratio between the integrated likelihoods $m_\gamma(\n y)/m_0(\n y)$
where
$$
m_\gamma(\n y)=\int M_\gamma(\n y\mid\n\beta_\gamma,\alpha,\sigma)\pi_\gamma(\n\beta_\gamma,\alpha,\sigma) d\n\beta_\gamma d\alpha d\sigma
$$
and $\pi_\gamma$ is the prior distribution, a quite delicate aspect of the Bayesian approach.

Most of the popular model selection priors in this context, like $g$-priors, Zellner-Siow priors, the hyper-$g$ priors, etc, share a similar functional form. This family of priors, that we call ``conventional'' priors, have been deeply studied by \citep{Baetal11} showing that they have many appealing theoretical properties. In this paper we propose an extension of the conventional priors that covers the situation with more possible regressors than data points and that has the original conventional priors as particular case. We call this priors regularized conventional priors. Our extension has important connections with other proposals in the literature like... We introduce the main motivating ideas in Section~\ref{motiva}. 

The rest of the paper is organized as follows.

\section{Conventional priors and motivating ideas} \label{motiva}
In this work, we adopt the term ``conventional'' \citep[used by][]{BerPer01} to refer to a big family of priors that are extremely popular in the literature. In the standard scenario with more data points than possible regressors ($n\ge p+k_0$), these are of the form $\pi_\gamma(\n\beta_\gamma,\alpha,\sigma)=\sigma^{-1}\pi_\gamma(\n\beta_\gamma\mid\alpha,\sigma)$, with the conditional distribution being an elliptical density of the type
\begin{equation}\label{conv}
\pi_\gamma(\n\beta_\gamma\mid\alpha,\sigma)=\int_0^\infty\, N_{k_\gamma}(\n\beta_\gamma\mid\n 0,t\n S_\gamma)\, p_n(t)\, dt,
\end{equation}
where $\n S_\gamma=\sigma^2[\n V_\gamma^t\n V_\gamma]^{-1}$ is the sampling variance matrix of the maximum likelihood estimator of $\n\beta_\gamma$ and $p_n(t)$ is a proper density that acts as a mixing density. The role of this matrix in $\n S_\gamma$ has been traditionally justified as giving sense to using the same improper prior distribution for common parameters $\pi(\alpha,\sigma)=\sigma^{-1}$ because this way common parameters are orthogonal to model specific parameters $\n\beta_\gamma$ (in an information Fisher's sense). Nevertheless, \cite{Baetal11} have shown that, in fact, using $\pi(\alpha,\sigma)=\sigma^{-1}$ can be formally justified with invariance and predictive matching argument and that orthogonality does not play any role. They have also shown that the use of $\n P_n$ in the definition of the prior scale is related with null predictive matching, providing a characterization of conventional priors. Interestingly, this criterion have important implications in this study as we will see.

Conventional priors is a big family of priors that contains, through a particular choice of $p_n$, very popular priors like the Zellner-Siow priors \citep{ZellSiow84}, the $g$-priors \citep{Zellner86, FLS01}, the hyper-$g$ priors \citep{liang08} or the robust priors 
\citep{Baetal11} just to mention some. Recently, \cite{Baetal11} have shown that conventional priors have optimal properties in the sense that they satisfy several formal criteria. In particular, \cite{Baetal11} showed that, irrespectively of $p_n$, conventional priors are measurement and group invariant and exact, dimensional and null predictive matching. 

It is also very convenient that conventional priors lead to simple expression for the Bayes factors:
\begin{equation}\label{cBF}
B_\gamma=\int (1+t\, Q_\gamma)^{-(n-k_0)/2}(1+t)^{(n-k_\gamma-k_0)/2}\, p_n(t)\, dt,
\end{equation}
where $Q_\gamma=SSE_\gamma/SSE_0$ is the ratio of sums of squared errors of $M_\gamma$ to $M_0$. One appealing characteristic of the Robust prior proposed in \cite{Baetal11} is that the above integral can be expressed in closed form using a Hypergeometric function. An alternative formula for the Bayes factor is
\begin{equation}\label{cBF2}
B_\gamma=\int (1+tn\, Q_\gamma)^{-(n-k_0)/2}(1+tn)^{(n-k_\gamma-k_0)/2}\, p_n^\star(t)\, dt,
\end{equation}
where $p_n^\star(t)=p_n(tn)n$.

As a direct consequence of next result, the matrix $\n S_\gamma$ is defined only when $n\ge k_\gamma+k_0$.
\begin{res}\label{rankS}
The rank of the $n\times k_\gamma$ matrix $\n V_\gamma$ is $n-k_0$ if $n< k_\gamma+k_0$ and $k_\gamma$ if $n\ge k_\gamma+k_0$.
\end{res}
\begin{proof} The case with $k_0=0$ is a straight consequence of the assumptions about the rank of $\n X_\gamma$. Show the case $k_0=1$
\end{proof}
The implication is that, when $n<p+k_0$, conventional priors are not defined for all competing models since models $M_\gamma$ with $k_\gamma+k_0>n$ would have an undefined prior scale matrix. Models in the model space can then be catalogued as ``regular'' (when $k_\gamma+k_0<n$), ``saturated'' (when $k_\gamma+k_0=n$) and ``singular'' (when $k_\gamma+k_0>n$).


In part because of the problem described above, the development of Bayesian methods when ${\cal M}$ contains singular models have been inspired by other sources of motivations different from the conventional tradition. Among these approaches highlight those based on regularization methods like Lasso (least absolute shrinkage and selection operator) introduced by \cite{Tib96} and their Bayesian counterparts \citep[see eg.][]{ParkCa08} of using a Laplace prior. The most appealing feature of Lasso is sparsity, meaning that when used over the full model ($\n\gamma=\n 1$) the estimation of certain regression parameters would be exactly zero. Hence, undoubtedly Lasso induces a type of variable selection but is not a formal model selection procedure since the most complex model is implicitly assumed as the true model (there is no model uncertainty considered). In practical terms, the immediate consequence is that there is no a way of measuring the uncertainty regarding the model selection exercise. That limitation has been noticed by \cite{Hans10,LyNtz13} who have embedded the Lasso approach into the formal framework of model selection adopting a multivariate Laplace prior for the specific regression parameters of each entertained model. Despite its undoubtedly value and interest, the only justification of these priors is their connection with the Lasso methodology and to the best of our knowledge there have not been proved any optimal property for these priors. The most distinctive feature of these priors with respect to the conventional priors is not the form of the prior itself (eg. the Laplace density can be defined as a mixture of a normal distribution) but the independency assumed among the regression parameters. This allows for a proper density but, as acknowledged by \cite{LyNtz13}, the Bayesian Lasso ``does not account for the structure of the covariates''. 

A compromise between considering the structure of covariates within a formal model selection problem are the ridge-inspired procedures by \cite{GupIbra07} and \cite{BaPo12}. These authors have incorporated dependence among the regression coefficients using an extension of the $g$-prior that circumvents the singularity of the conventional scale matrix through the introduction of a ridge parameter $\lambda$. In particular, they propose using the scale matrix $\sigma^2[\n V_\gamma^t\n V_\gamma+\lambda\n I]^{-1}$ (normally after a transformation of the covariates so that they have unitary scale). A drawback in this setting is the specification of the ridge parameter  which in principle may have a strong impact on results. Also is that the priors used for the regular models are not conventional priors and do not share the optimal properties of the conventional priors. 

Another more sophisticated and also interesting extension of conventional priors is \cite{MaGe12} that handles the case where $n\le p$ through a singular value decomposition of $\n X_\gamma$. (for our records: this is not really an extension as it does not have the $g$ as a particular case hence the optimal properties are not inherited. For instance, is their prior invariant under changes in the units?).

In Section~\ref{nltp} we define a generalization of conventional priors that we call regularized conventional priors. These are proper priors and are based on using, for the conditional scale for $\n\beta_\gamma$ a {\em non-singular} generalized inverse matrix of $\n V_\gamma^t\n V_\gamma$. Obviously, such matrices equal the inverse of $\n V_\gamma^t\n V_\gamma$ for regular models, justifying that our proposal generalizes the conventional priors. The form of the scale matrix has connections with the ridge-based approaches by \cite{GupIbra07} and \cite{BaPo12}. For singular models, regularized conventional priors are not univocally defined, but within a model, the posterior distribution of any estimable function is unique. This paper is about model selection and we  will show a surprising result that, for singular models, the associated Bayes factor is one. As we will see this can be viewed as an extension of the null predictive matching criterion and is also congruent with full rank factorizations of the problem. The impossibility of distinction between singular models $M_\gamma$ is also implicitly present in other methodologies like lasso, where the result can never be one of such models (Rosset and Zhu, 2007 copy reference at the end!).

Work more on this paragraph: states the important conclusion that then, what remains is nothing more than a multiple testing problem. 
When $p$ is much larger than $n$ there are a huge number of regular models that what arises is a multiple testing problem and a control for multiplicity is called for. The discussion and arguments in  \cite{ScottBerger06} points out that the proper Bayesian way of handling multiplicity issues is through the prior distribution over the model space. We adopt here their recommendation of using the prior
\begin{equation}\label{SB}
P(M_\gamma)=1/(p+1){p \choose k_\gamma}.
\end{equation}
which penalizes models in dimensions containing a large number models, where it is precisely more likely to appear more `false signals'.

The rest of the paper is organized as follows.

\section{Regularized conventional priors}\label{nltp}
The non-singularity of the scale matrix in the ridge-based proposals by \cite{GupIbra07} and \cite{BaPo12} is due to the addition of the diagonal dominant matrix $\lambda\n I$ which intuitively results in a substantial modification of the original scale matrix. This modification seems unneeded for regular models where $\n S_\gamma$ is non-singular and here we study alternative ways to define the scale matrix.

In the following definition ${\cal R}(\cdot)$ denotes the sub-space spanned by the rows of the matrix in the argument.

\begin{defn}
For $M_\gamma$, with parameters $(\alpha,\n\beta_\gamma,\sigma^2)$ the regularized conventional prior for $\n\beta_\gamma\mid \alpha,\sigma^2$ is any proper prior of the form
\begin{equation}\label{Rconv}
\pi_\gamma(\n\beta_\gamma\mid\alpha,\sigma)=\int_0^\infty\, N_{k_\gamma}(\n\beta_\gamma\mid\n 0,t\n S_\gamma^\star)\, p_n(t)\, dt,
\end{equation}
where $p_n(t)$ is a mixing density; $\n S_\gamma^\star=\sigma^2[\n V_\gamma^t \n V_\gamma+\n T_\gamma]^{-1}$ and $\n T_\gamma$ any symmetric semi positive definite matrix such that ${\cal{R}}(\n V_\gamma^t \n V_\gamma)\oplus \cal{R}(\n T_\gamma)={\mathbb{R}}$$^{k_\gamma}$.
\end{defn}

Note that these priors extend the conventional priors since for regular models, ${\cal{R}}(\n V_\gamma^t \n V_\gamma)=\mathbb{R}$$^{k_\gamma}$ and by definition $\n T_\gamma$ is the zero matrix. Further, for singular models these priors are proper and hence valid for model selection. Finally, note that these priors exist since we can always take as $\n T_\gamma=\n C_\gamma^t\n C_\gamma$ where $\n C_\gamma:(k_\gamma-n+k_0)\times k_\gamma$ is of full rank and with rows that are independent of the rows of $\n V_\gamma$. These way, for singular models, the regularized conventional priors can be seen as conventional priors but with respect to the `extended' design matrix
$$
\Big(\begin{array}{c}\n V_\gamma \\ \n C_\gamma\end{array}\Big),
$$ 
on which the `missed' values of the covariates ($k_\gamma-n+k_0$ cases up to completing the $k_\gamma$) are replaced by the imputed values in $\n C_\gamma$. 

Of course, the regularized conventional priors depend on $\n T_\gamma$ (or on the imputed values $\n C_\gamma$), nevertheless the dependence happens in a way that the corresponding prior does not influence the likelihood. The reason why this is the case is contained in the following important result.

\begin{res} Let $\n T_\gamma$ a matrix defined as in \ref{Rconv}, then the regular matrix $[\n V_\gamma^t \n V_\gamma+\n T_\gamma]^{-1}$ is a generalized inverse of $\n V_\gamma^t \n V_\gamma$.
\end{res}
\begin{proof} An immediate consequence of Theorem 18.2.5, page 421 (Harville, 1997) and the fact that $[\n V_\gamma^t \n V_\gamma+\n T_\gamma]^{-1}$ is a generalized inverse (because it is an inverse) of $[\n V_\gamma^t \n V_\gamma+\n T_\gamma]$.
\end{proof}

The above result justifies a very appealing interpretation of regularized conventional priors since the (not defined for all models) matrix $(\n V_\gamma^t \n V_\gamma)^{-1}$ is replaced by a {\em regular} generalized inverse $(\n V_\gamma^t \n V_\gamma)^{-}$. Said other way, we can keep the desired scale (that based on $(\n V_\gamma^t \n V_\gamma)$) but still using a regular matrix, hence defining a valid model selection prior.

Nevertheless, the regularized conventional priors are not unique (as there are many different ways of defining the matrix $\n T_\gamma$). Nevertheless a crucial property of these priors is invariance under such choice when estimating, given a model, a parameter univocally informed by the data (what are usually called estimable functions, see Harville 19??). We take this as evidence that regularized conventional priors are sensible priors and hence satisfy the principle in Berger and Pericchi (2001) about model selection priors.

\begin{res}
Let $M_\gamma$ be any singular model and let $\n\theta$ be an estimable function, then the posterior distribution of $\n\theta$ given that $M_\gamma$ is the true does not depend on the choice of the matrix $\n T_\gamma$.
\end{res}
\begin{proof}(Removing the subindex $\gamma$ for simplicity)
First show that
$$
\n\beta_\mid \alpha,\sigma,t,\n y\sim N_k(\n m,\n\Sigma)
$$
where 
$$
\n m=[\n V^t \n V(1+t^{-1})+\n T t^{-1}]^{-1}\n V^t \n y
$$
and
$$
\n\Sigma=\sigma^2[\n V^t \n V(1+t^{-1})+\n T t^{-1}]^{-1}.
$$
Secondly show that $[\n V^t \n V(1+t^{-1})+\n T t^{-1}]^{-1}$ is a generalized inverse of $\n V^t \n V(1+t^{-1})$. It is well known that the matrix product $\n V[\n V^t \n V(1+t^{-1})]^-\n V^t$ does not depend on the generalized inverse implying that in our case
$$
\n H=\n V[\n V^t \n V(1+t^{-1})+\n T t^{-1}]^{-1}\n V^t
$$
 particularly does not depend on $\n T$. 

Finally, notice that estimable functions are of the type $\n\theta=\n t^t \n V$ where $\n t$ is any vector of conformably dimension with posterior distribution
$$
\n\theta_\mid \alpha,\sigma,t,\n y\sim N_k(\n t^t \n H \n y, \n t^t \n H \n t),
$$
which is independent on $\n T$.
\end{proof}
(Perhaps show that the lasso and ridge-based approaches depends on several choices?)

Our aim was on model selection. In the next result we show that the resulting Bayes factors of any singular model to the null is one.

\begin{res}\label{uBF} 
Let $M_\gamma$ be a singular model, then $B_\gamma=1$ independently of the choice of $\n T_\gamma$ and of $p_n()$.
\end{res}
\begin{proof}
(Removing the subindex $\gamma$ for simplicity)
According to Result~\ref{rankS}, if $M$ is singular then $rank(\n V)=n-k_0$ and this matrix admits a full rank factorization of the type $\n V=\n L\n R$ where $\n L:n\times (n-k_0)$ and $\n R:(n-k_0)\times k$ are of full rank. Now in computing the integral $m_\gamma(\n y)$ apply the change of variables $\n\beta_R=\n R\n\beta$ and $\n\beta_C=\n C \n\beta$, (here $\n C:(k-n+k_0)\times k$ is of full rank such that $\n T=\n C^t\n C$) to show that
$$
m_\gamma(\n y)=m_0(\n y)\, 	det(\n L^t(\n I-\n P_n)\n L)^{-1/2}\, det(\n V^t\n V+\n T)^{1/2}\, det\Big(\begin{array}{c}\n R  \\ \n C \end{array}\Big)^{-1},
$$
(mainly due to the null predictive matching property of conventional priors). Now show that the factor to the right of $m_0(\n y)$ above is 1.
\end{proof}

\section{Unitary Bayes factors}
The main consequence of adopting the regularized conventional priors in a variable selection problem containing singular models is that for singular models $B_\gamma=1$ and for regular models $B_\gamma$ is a standard conventional Bayes factor.. 

There are a number of independent arguments that also support unitary Bayes factors for singular models.

\paragraph{Null predictive matching} This is one of the criteria proposed by \cite{Baetal11} to construct sensible objective prior distributions and reflects the idea -starting with \cite{Jef61}- that data of minimal size for a given model should not allow one to distinguish between that model and the null. 
In the regular case this implies that, when $n=k_\gamma+k_0$ (for saturated models) the Bayes factor of $M_\gamma$ to the null must be one. Interestingly, as highlighted by \cite{Baetal11}, null predictive matching provides a characterization for the scale $\n S_\gamma$ since no other matrix (or a multiple) can achieve this predictive matching.
This was taken by \cite{Baetal11} as positive evidence in favour of conventional priors (the ridge-based proposals do not have this property).

The situation with $n< k_\gamma+k_0$ is an extreme case of data of minimal size (more parameters than observations was explicitly mentioned by \cite{Jef61}: check!) in the sense that like in the saturated case there is only enough data as to estimate estimable functions in $M_\gamma$ but not to distinguish it from then null.

\paragraph{Reparameterization} Consider first the following result:

\begin{res}
Any singular model $M_\gamma$ admits a reparameterization as a saturated model $M_\gamma^\star$ (ie with $k_\gamma^\star=n-k_0$). 	
\end{res}
\begin{proof} (Revise to contain the case $k_0=0$)
Without loss of generality any model $M_\gamma$ can be reparameterized as
$$
M_\gamma: \,\, \n y=\alpha\n 1_n+\n V_\gamma\n\beta_\gamma+\n\epsilon,\,\,\,\n\epsilon\sim N(\n 0,\sigma^2\n I_n),
$$
where $\n V_\gamma=(\n I-\n P_0)\n X_\gamma$. According to Result~\ref{rankS}, if $M_\gamma$ is singular then $rank(\n V_\gamma)=n-1$ and this matrix admits a full rank factorization of the type $\n V_\gamma=\n L_\gamma\n R_\gamma$ where $\n L_\gamma:n\times (n-1)$ is of full column rank and $\n R_\gamma:(n-1)\times k_\gamma$. Hence we can parameterize $M_\gamma$ as
\begin{equation}\label{rep}
M_\gamma^\star:\,\, \n y=\alpha\n 1_n+\n L_\gamma\n\beta_\gamma^\star+\n\epsilon,\,\,\,\n\epsilon\sim N(\n 0,\sigma^2\n I_n),
\end{equation}
where $\n\beta_\gamma^\star=\n R_\gamma\n\beta_\gamma$ is now $n-1$-dimensional. Show that $M_\gamma^\star$ can be constructed in a way that it is a saturated (original) model.
\end{proof}

According to the result above, singular models $M_\gamma$ have an equivalent representations as saturated models (which, recall have associated a conventional Bayes factor to the null of one). This coincidence does not depend on the arbitrary choice of the matrices $\n L$ and $\n R$ used to construct the reparameterization. Notice that, a prior distribution without the null predictive matching property (out of conventional family) would easily lead to a Bayes factor that depends on these matrices.

In what follows in this document we use the conventional Robust prior in \cite{Baetal11} because it is our preferred prior, but everything applies to the family of conventional priors.

\section{The posterior distribution when $p>>n$}
Up to here the conclusion is that there is no informative content (coming from the data) in singular models (of course this does not imply that they do not influence the posterior distribution). Since $n$ is expected to be much bigger than $n$, this implies that there is only information in a very few number of models (in the example in Section~\ref{sec.example} with $p=8408$ and $n=41$ the proportion of regular models over the total number of models is of the order $10^{-2000}$). The obvious and crucial question that arises is if we can learn something or not.

In what follows we denote ${\cal M}^S$ the set of singular or saturated models (these share a unitary Bayes factor making it convenient to group in a common set) and denote ${\cal M}^R$ the rest (formed by the regular models). See Table~\ref{TM}.

\begin{table}[t!]
\begin{center}
{\small\scalebox{0.85}{
\begin{tabular}{clc}
Dimension & Type & Subsets of ${\cal M}$\\
\hline
$k_\gamma+k_0<n$ & Regular & ${\cal M}^R$\\
$k_\gamma+k_0=n$ & Saturated & ${\cal M}^S$\\
$k_\gamma+k_0>n$ & Singular & ${\cal M}^S$\\
\hline
\end{tabular}
}}
\end{center}\caption{Type of models}\label{TM}
\end{table}

Application of Bayes' theorem leads that the posterior distribution can be expressed as a weighted average over the above defined subsets of ${\cal M}$ weighted by their corresponding posterior probabilities that we denote (recall $M^T$ represents the true model)
$$
P^S=Pr(M^T\in{\cal M}^S\mid\n y)
$$ 
and $(1-P^S)$. 
Notice that
$$
P^S=\frac{\sum_{M_\gamma\in{\cal M}^S}B_\gamma\, Pr(M_\gamma)}{\sum_{M_\gamma\in{\cal M}^S}B_\gamma\, Pr(M_\gamma)+\sum_{M_\gamma\in{\cal M}^R}B_\gamma\, Pr(M_\gamma)}
$$
then, and because of $B_\gamma=1$ for $M_\gamma\in{\cal M}^S$:
$$
P^S=\frac{Pr({\cal M}^S)}{Pr({\cal M}^S)+Pr({\cal M}^R)\,C(n,p)}
$$
where $C(n,p)$ is the normalizing constant conditionally on $M^T\in{\cal M}^R$, that is,
$$
C(n,p)=\sum_{M_\gamma\in{\cal M}^R}B_\gamma\, Pr(M_\gamma\mid M^T\in{\cal M}^R).
$$
For the particular case of the prior in (\ref{SB}), notice that of the $p+1$ different dimensions, $n-k_0$ correspond to ${\cal M}^R$ and the rest, $p-n+k_0+1$ belong to ${\cal M}^S$ and then
$$
Pr({\cal M}^S)=(p-n+k_0+1)/(p+1),
$$
and hence
\begin{equation}\label{pS}
P^S=\frac{p-n+k_0+1}{p-n+k_0+1+(n-k_0)\, C(n,p)}.
\end{equation}
Now, any summary of the posterior distribution can be constructed as weighted averages. One popular of such are the inclusion probabilities. The posterior inclusion probability of $x_i$ is
\begin{equation}\label{qi}
q_i=Pr(x_i\in M^T\mid\n y)=q_i^R\,(1-P^S)+q_i^S\,P^S
\end{equation}
where $q_i^R$ denotes the inclusion probability conditional on $M^T\in{\cal M}^R$ and identical notation for $q_i^S$. For the case of the prior in (\ref{SB}) it can be seen that
$$
q_i^S=\frac{1}{2}\frac{p(p+1)-(n-k_0)(n-k_0-1)}{p(p-n+k_0+1)}
$$
that tends to 1/2 when $p$ grows and $n$ is either constant or grows at a rate $n\approx p^a$, for some $0<a< 1$ or $n\approx \log(p)$ (here $\approx$ means asymptotic equivalence). When $n$ grows linearly with $p$, say $n\approx f\,p$ for some $0<f<1$ then $q_i^S$ tends to $(1-f^2)/2$.

Then, if $p$ is large enough, the informative content in $q_i$ depends on the magnitude of $P^S$. 

\section{The methodology in practice and an illustrative application}\label{sec.example}
When ${\cal M}$ is very large ($p$ is in the hundreds or more) it is quite difficult to figure out an algorithm that convincingly explores the model space. Nevertheless, for the problem here analyzed with $n<<p$, we have argued that we know what happens in ${\cal M}^S$, a huge subset of the whole model space. The challenge is then how to manage that information to produce reliable results. 

In the previous section we have seen that what is essentially needed is an estimation of $P^S$ and $q_i^R$, and both quantities can be estimated with methods exploring efficiently ${\cal M}^R$ (still a moderate to large model space). We put in practice the following scheme:

\begin{enumerate}
	\item Use the Gibbs sampling algorithm studied and recommended in \cite{Ga-DoMa-Be13} to obtain two samples of size $N$ of the posterior distribution over ${\cal M}^R$ (i.e. $Pr(M_\gamma)\propto$(\ref{SB}) if $M_\gamma\in{\cal M}^R$ and zero otherwise). 
	
	\item Check that convergence has been achieved (discharge burnin samples if needed) and use eq.(35) in \cite{GeorgeMcCulloch97} to estimate the normalizing constant $C(n,p)$ (August 15: despite notation, this $C$ is not a normalizing constant). Use it, in combination with (\ref{pS}) to estimate $P^S$. At this point you know who wins.
	
	\item Combine both samples to estimate $q_i^R$ and use formula in (\ref{qi}) and $P^S$ to compute an estimate of $q_i$.
\end{enumerate}

All steps above described can be done with a small $p>n$ modification of the {\tt R} package {\tt BayesVarSel} by \cite{Ga-DoFor12} available upon requests from the authors.

We illustrate the methodology using the simulation study based on a real dataset in \cite{Hansetal07}. These data consist on a gene expression dataset from a survival study in brain cancer (add reference) with $n$ patients and $p=8408$ genes from a tumor specimen. Exactly as in \cite{Hansetal07} we define the `true' data generating model as
\begin{equation}\label{sim.truth}
y_i=1.3x_{i1}+.3x_{i2}-1.2x_{i3}-.5x_{i4}+\epsilon_i,
\end{equation}
where $\epsilon_i\sim N(0, 0.5)$ from which we simulated one dataset with $n=41$ observations. As it is described in that paper, these four covariates where chosen in part because of previous information about these genes and also because they exhibit some correlation with other genes in the dataset. (Important: this problem does not have intercept and formulas have to be re-written for this situation. Results here presented already takes into account this). We run the algorithm for $N=11000$ iterations, of which the first 1000 were discharged. Results are summarized in Table~\ref{Shotgun.Tab}.

\begin{table}[t!]
\begin{center}
{\small\scalebox{0.85}{
\begin{tabular}{ccccccccc}
n & $P^S$ & $q_1$ & $q_2$ & $q_3$ & $q_4$ & $\bar{q}_{-T}$ & $q^U_{-T}$ & \mbox{HPM}\\
\hline
  {\bf 41} & 0.004 & 0.843 & 0.154 & 0.766 & 0.002 & 0.002 & 0.038 & $\{x_1,x_3\}$\\[.25cm]
  30 & 0.320 & 0.300 & 0.171 & 0.173 & 0.160 & 0.159 & 0.251 & $\{x_1,x_3\}$\\
  20 & 0.830 & 0.417 & 0.416 & 0.415 & 0.415 & 0.414 & 0.419 & $\{x_{4026},x_{7748}\}$\\
  10 & 0.995 & 0.497 & 0.497 & 0.497 & 0.497 & 0.497 & 0.497 & \{Null,Full\}\\
\hline
\end{tabular}
}}
\end{center}
\caption{\small \cite{Hansetal07} dataset. Keys: $q_i$ is the inclusion probability for $x_i$ ($i=1,2,3,4$) and $\bar{q}_{-T}$, $q^U_{-T}$ are 
respectively the mean and maximum of the inclusion probabilities for the spurious variables. HPM is the estimated most probable a posteriori model.
}
\label{Shotgun.Tab}
\end{table}

First observation is that the posterior probability of the singular subset is very small (0.004) and basically the posterior distribution concentrates in the regular part. The results are quite informative: two of the four `true' covariates ($x_1$ and $x_3$) have a large inclusion probability and none of the 8404 spurious covariates have a non-negligible probability (the upper bound was 0.038). The variable $x_2$ has a small inclusion probability (0.154) but that is at least four times any inclusion probability of the spurious covariates. Interestingly, the highest posterior probability model (HPM), $\{x_1,x_3\}$ also gives extra evidence about the importance of these covariates in the experiment. 

\begin{table}
\begin{center}
\begin{tabular}{|c|c| c c c c|c|c}
  \hline
  n & $p^s$ & $q_1$ & $q_2$ & $q_3$ & $q_4$ & $q_{mean}$ & $q^U_{-T}$\\
  \hline
  41 (beta(1,f(p))) & 2.91e-6 & 0.809 & 0.156 & 0.726 & 5.45e-4 & 3.5e-4 & 0.028 \\
  41 (beta(1,99)) & 2.53e-5 & 0.805 & 0.151 & 0.725 & 0.001 & 3.6e-4 & 0.034 \\
  41 (beta(1,9)) & 0.003 & 0.784 & 0.153 & 0.704 & 0.000 & 4.3e-4 & 0.045 \\
  41 & 0.004 & 0.843 & 0.154 & 0.766 & 0.002 & 0.002 & 0.038 \\
  30 & 0.320 & 0.300 & 0.171 & 0.173 & 0.160 & 0.159 & 0.251 \\
  20 & 0.830 & 0.417 & 0.416 & 0.415 & 0.415 & 0.414 & 0.419 \\
  10 & 0.995 & 0.500 & 0.500 & 0.500 & 0.500 & 0.500 & 0.500 \\
  \hline
\end{tabular}
\caption{}
\label{}
\end{center}
\end{table}

Finally, to analyze the impact of $n$ over $p$, we repeated the experiment with the first 10, 20 and 30 observations. Results are included also in Table~\ref{Shotgun.Tab}. There we can clearly seen how the informative content in the data is overwhelmed by a large $p$ when $n$ is small. In the extreme, when $n=10$, $P^S$ is 0.995 and the posterior and prior distributions basically coincide. When $n=30$ we have $P^S=0.320$ and the information in the data starts being relevant and what we see there is the parsimony of the Bayesian approach and all the variables have a small inclusion probability. Still the HPM points to the importance of $\{x_1,x_3\}$.

\begin{figure}[!t]
\begin{center}
{\small\scalebox{0.75}{
\begin{tabular}{cc}
\includegraphics[width=205pt,height=200pt]{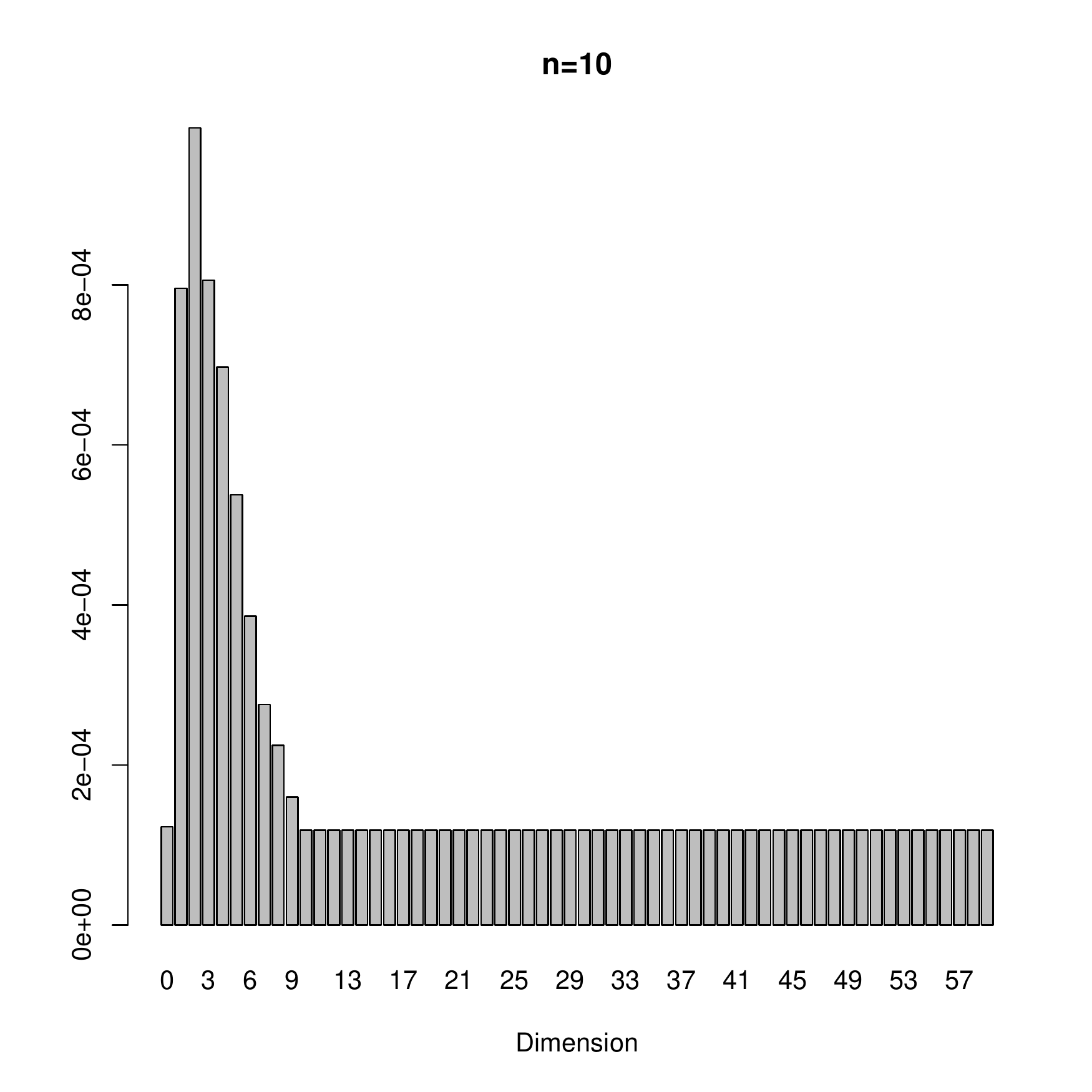} &
\includegraphics[width=205pt,height=200pt]{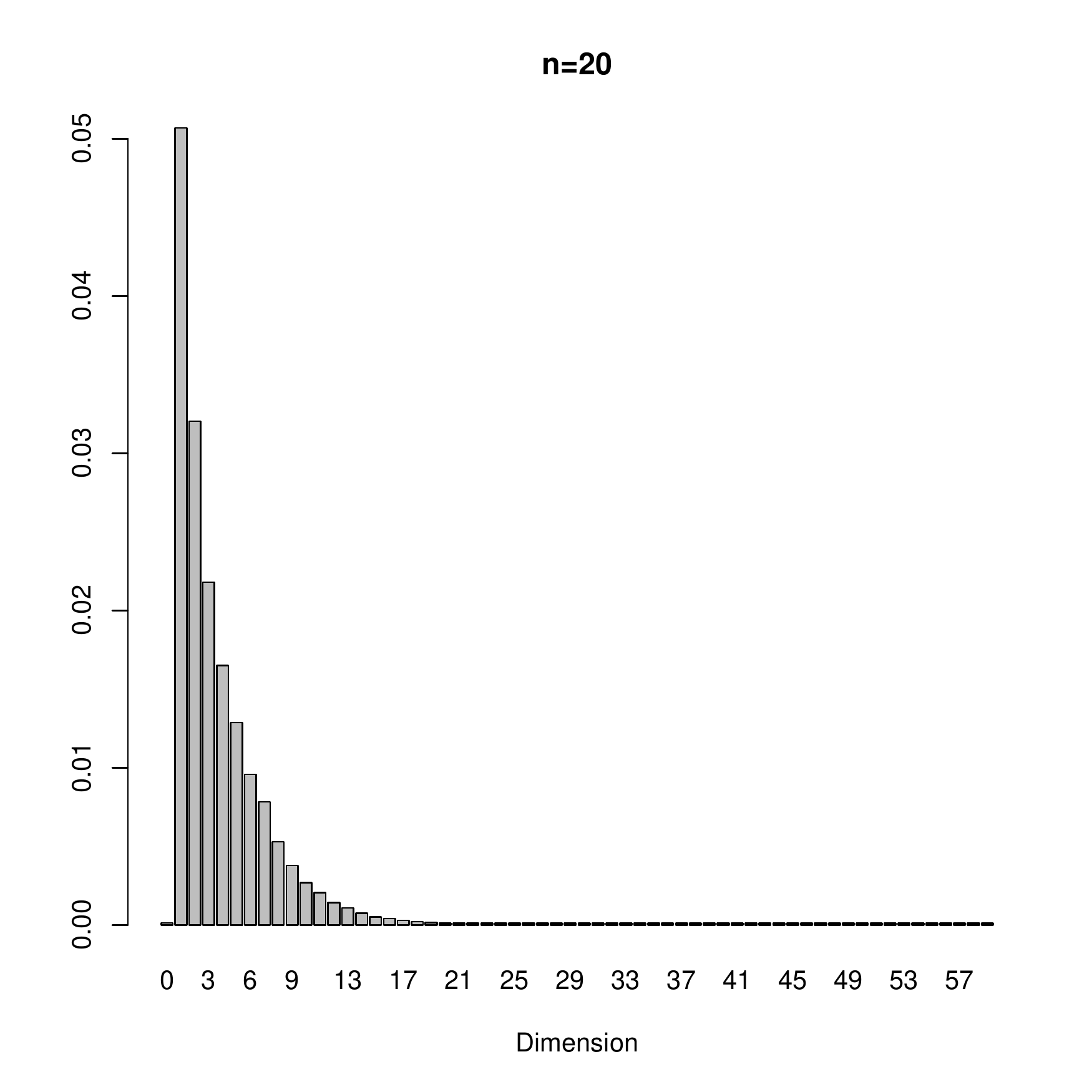}\\
\includegraphics[width=205pt,height=200pt]{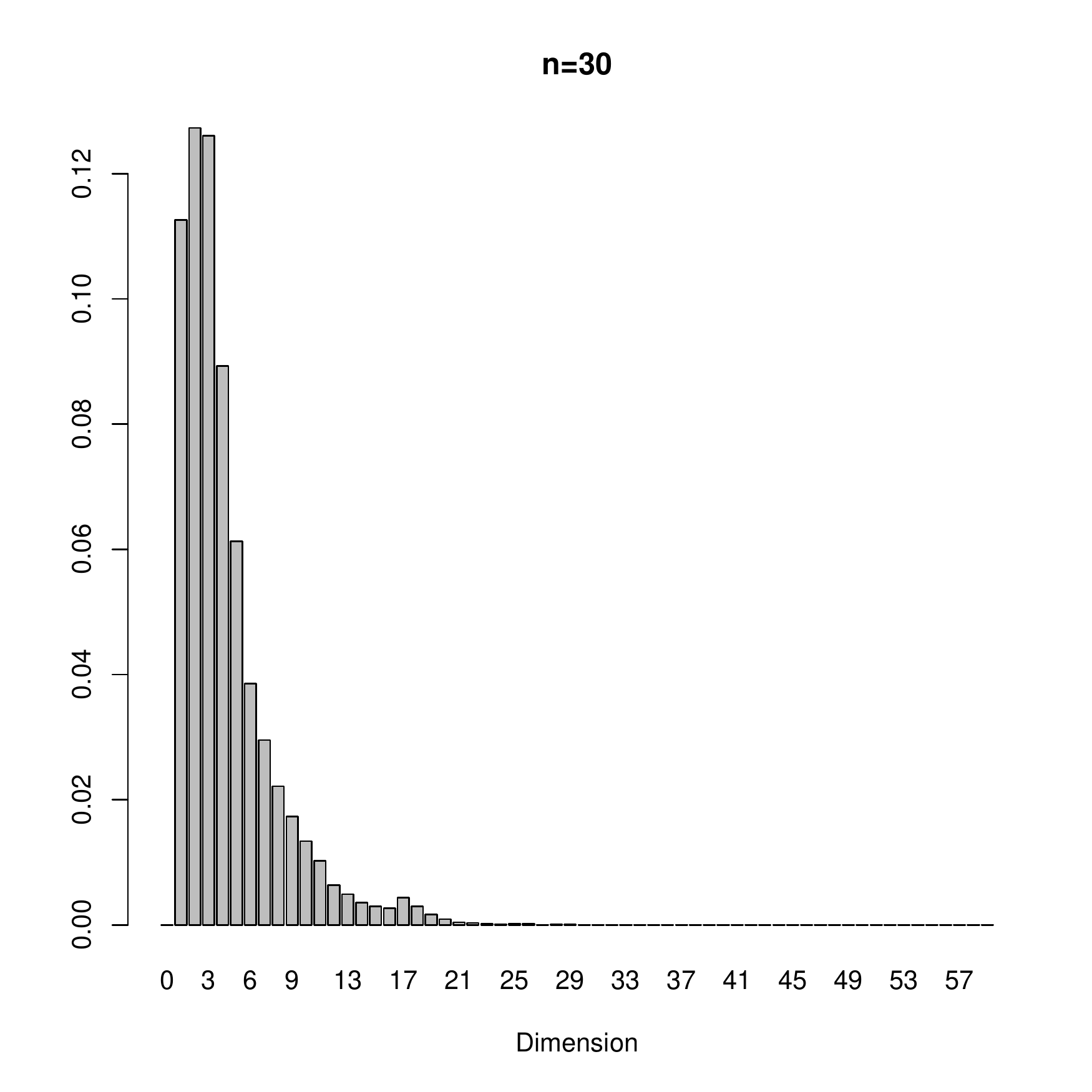} &
\includegraphics[width=205pt,height=200pt]{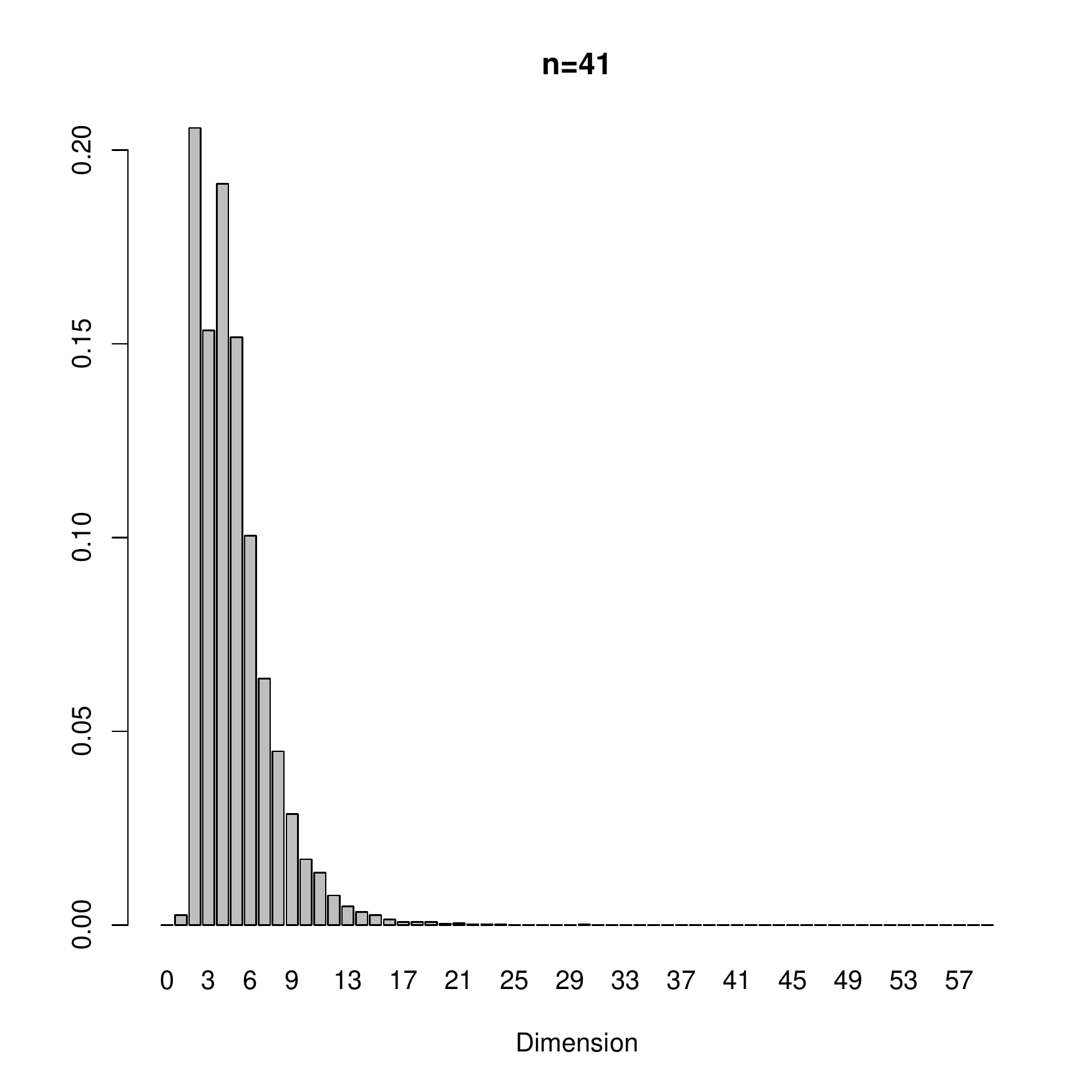}\\

\end{tabular}
}}
\caption{{\small Posterior probabilities of the dimension of the true model (in the x-axis only represented the first 60 values).}}
\label{ShotGun}
\end{center}
\end{figure}

Also, in Figure~\ref{ShotGun} we have represented the posterior distribution of the dimension of the true model for the different $n$'s. What do we learn?

\begin{figure}[!t]
\begin{center}
{\small\scalebox{0.75}{
\includegraphics[width=205pt,height=200pt]{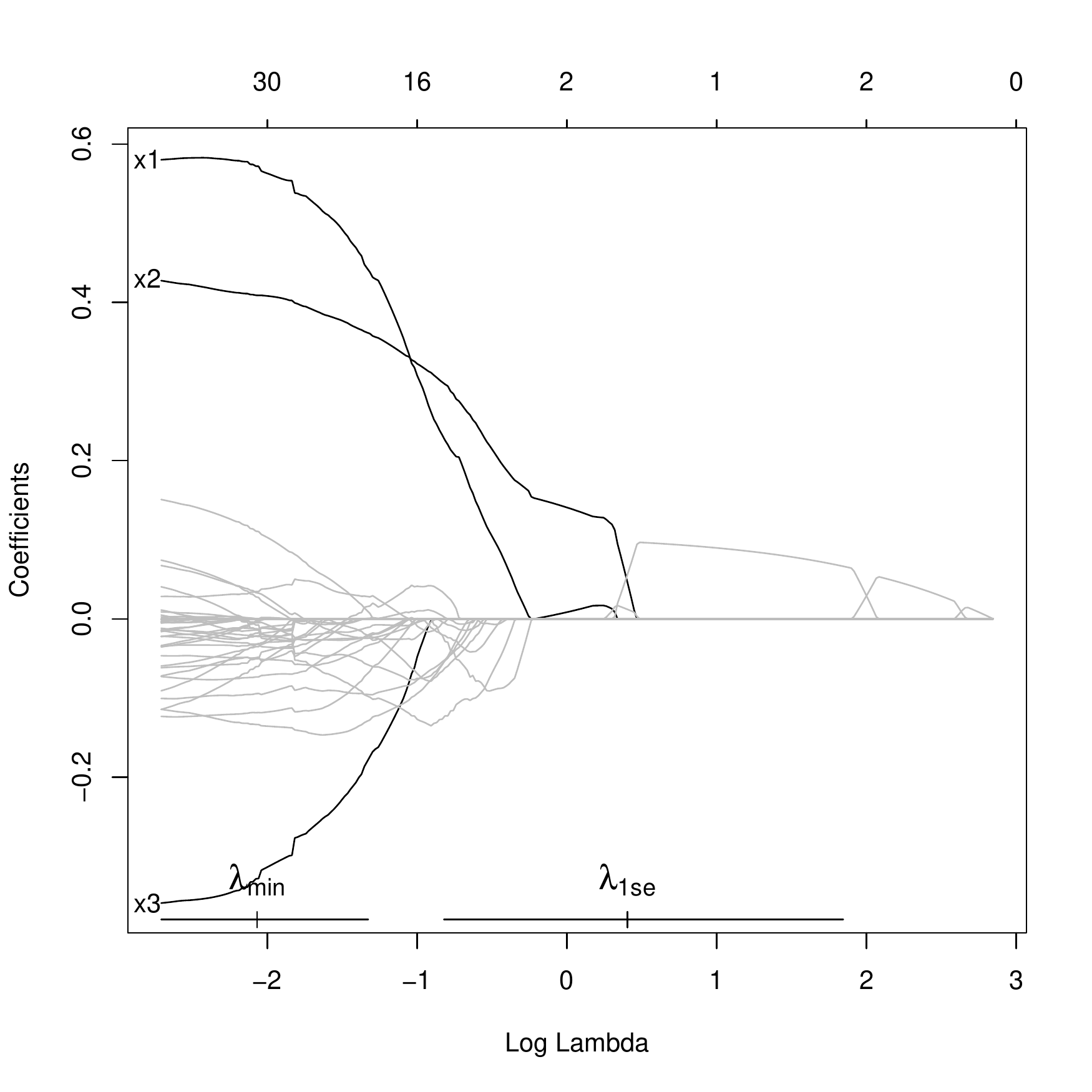}
}
}
\caption{{\small Lasso applied to the example. as a function of (log-)$\lambda$ --the penalisation parameter on the l1-norm of the coefficients of the variables included in the model-- the estimated values of such coefficients. The upper horizontal axis shows for some values of (log-)$\lambda$ the number of variables included in the model. Black lines correspond to the values of the coefficients for the `true' explanatory variables ($x_1$ to $x_4$) meanwhile gray lines stand for spurious variable included in the model for the different values of $\lambda$.}}
\label{LassoSG}
\end{center}
\end{figure}

Although Lasso's and our approach's results are hardly comparable (one depends on a penalty, the other one summarises its results in terms of posterior probabilities, ...), the popularity of Lasso for variables selection in the $n<<p$ setting worths a comparison with our proposal. Namely, we have run Lasso in our data set by using {\tt glmnet}, the {\tt R} package by \cite{Frietal10}. Figure~\ref{LassoSG} shows the Lasso fit for our whole dataset ($n=41$ observations). Note that $x_4$ does not get into the Lasso fit for any value of $\lambda$, meanwhile $x_3$ is included into the model for a few values of log-$\lambda$ around 0 and after being removed it is included again in models with lower values of log-$\lambda$ (lower than -1).

The value of $\lambda$ to be used in Lasso is often chosen by crossvalidation. Namely, two criteria are particularly popular for setting it: the value yielding the minimum cross-validated error (MSE in our case), that we call $\lambda_{min}$, and  the `one-standard error rule' i.e. that value of $\lambda$ yielding the simplest model less than one standard error away of $\lambda_{min}$. We call  this last criterion $\lambda_{1se}$ wich is intended to be a parsimonious alternative to $\lambda_{min}$. The lower side of Figure~\ref{LassoSG} show the estimated values for $\lambda_{min}$ and $\lambda_{1se}$. Since these values are chosen by cross-validation they yield different values for different runs of the cross-validation thus we have plotted for each of them a segment covering the central 80\% of the values obtained in 100 different cross-validations and the median value achieved in all those runs. 

According to $\lambda_{min}$ $x_1$, $x_2$ and $x_3$ are selected plus 17 to 38 spurious variables while $\lambda_{1se}$ selects  models ranging from 1 to 13 variables (the median value of $\lambda_{1se}$ would select $\{x_2, x_{3780}, x_{4494}\}$ so the results of the  $\lambda_{1se}$ seem more sensible than those derived from  $\lambda_{1se}$.

\bibliography{$HOME/Mywork/bibliografia/mibibliografia,$HOME/Mywork/bibliografia/Anabel,$HOME/Mywork/bibliografia/MSComputation}

\bibliographystyle{plainnat}

\end{document}